# On Topological Kondo Insulator Plasmonics

*Partha Goswami*

*D.B.College, University of Delhi, Kalkaji, New Delhi-110019, India*
*physicsgoswami@gmail.com*

**Abstract**. We study surface state plasmonics in the periodic Anderson model of a bulk topological Kondo insulator (TKI) system, involving strongly correlated $f$ orbital electrons screened by weakly correlated $d$ orbital electrons, performing a mean-field theoretic calculation within the frame-work of slave-boson protocol. We include the effect of magnetic impurities as well which do not act as scattering agent in our scheme; the exchange interaction due to them is included in the band dispersion. The interaction opens a gap at the surface states with gapless Dirac dispersion. We find that bulk insulation together with the surface metallicity is possible for the system albeit only under the condition demanding strong $f$-electron localization. The possibility of the intra-band and the inter-band acoustic plasmons for the gapped surface states exists in our report. The dominance of the bulk metallicity and the strong incarceration capability of these plasmons are other notable outcome of the present work.

## 1. Introduction

The Periodic Anderson model (PAM) **[1-6, 25]** basically involves highly correlated electrons (localized magnetic moments) in one $f$ orbital which are screened by weakly correlated electrons in a second $d$ orbital (a broad band of width ,say,W). Although this model has been thoroughly examined for several decades, the model itself and its extensions **[5,6]** are still relevant for the theoretical condensed-matter physics. A captivating development in recent years is the discovery that Kondo insulators can develop topological order to form a topological Kondo insulator (TKI). The theoretical description of a Kondo insulator **[7-10, 16]** is usually based on PAM. In this paper we shall focus on a generic topological Kondo insulator with the simplest band structure and the associated plasmonics. The plasmons are defined as longitudinal in-phase oscillation of all the carriers driven by the self-consistent electric field generated by the local variation in charge density. We find the intra-band and the inter-band acoustic plasmons for the gapped surface states with strong incarceration capability.

It may be mentioned that the first example of correlated topological Kondo insulator(TKI) **[1,7-10]** is $SmB_6$. It has a cubic crystal structure of the cesium-chloride type with a lattice constant of $a \approx 4.13$ Å. At high temperature, the material behaves like a metal, but when reducing the temperature below ∼ 50 K, it exhibits insulating behavior **[11]**. The hybridization gap (Δ) has been measured at ∼15 meV **[11,12]**. This constitutes typical behaviour for a Kondo insulator. However, $SmB_6$ shows some unusual properties in addition. It has been found **[13]** that the resistivity of $SmB_6$ increases like an insulator but saturates at temperatures below 5K upon decreasing the temperature. While non-magnetic impurities do not influence the saturation anyway, doping $SmB_6$ with magnetic impurities does **[14,15].** The unusual properties of this rare earth hexa-borides are ascribed to the interaction of the 5$d$ and 4$f$ electrons of rare earth element with the 2$p$ conduction electrons of Boron. Unlike the well-known topological insulators, such as $Bi_2Se_3$ family of materials, in TKIs the strong correlation effects among the 4$f$-electrons are very important. It brings about the strong modification of the 4$f$ -band width. It may be mentioned that in the $Bi_2Se_3$ family of materials, the band inversion happens between two bands both with the $p$ character and similar band width. The situation in $SmB_6$ is quite different, where the band inversion happens between 5$d$ and 4$f$ bands with the band widths differing by several orders of magnitude, which leads to very unique low energy electronic structure.

To investigate the bulk model, the slave boson technique **[1,17-20]** is usually employed. The slave particle protocol is based on the assumption of spin-charge separation in the strongly correlated electron systems in Mott insulators. The surmise is that electrons can metamorphose into spinons and chargons. But to preserve the fermion statistics of the electrons, the spinon-chargon bound state must be fermionic, so the simplest way is to ascribe the fermion statistics to one of them: if the spinon is fermionic then the chargon should be bosonic (slave-boson), or if the chargon is fermionic then the spinon should be bosonic (slave-fermion). The two approaches are just two low-energy effective theories of the complete-fractionalization theory **[21,22].** We use the slave-particle mean-field theory (MFT) **[1,12-14]** which allows the study of low-energy regime of a Kondo system with a quadratic single-particle Hamiltonian in the limit of the *f*-electron correlations being larger than all other energy scales in the problem. We obtain the self-consistent equations for MFT parameters minimizing the grand canonical potential of the system with relative to these parameters as in ref. **[1]**. The parameters enforce constraints on the pseudo-particles due to the infinite Coulomb repulsion and the need of the formation of singlet states between an itinerant electron and a localized fermion at each lattice site in order to have a Kondo insulator. The technique used by us to calculate various thermodynamic averages is the Green's function method as was used by Legner **[1]**. The theory allows to conveniently circumvent complications associated with formally, infinite repulsion between the *f*-electrons by "splitting" the physical *f*-electron into a product of a fermion and a slave boson, supplemented with a constraint to remove the double occupancy. The temperature range that slave-particle mean-field-theory is valid, limited to very low values ($T \to 0$). Furthermore, MFT involves condensing the boson field and neglecting all of its dynamics. This effectively leads to a non-interacting model of an insulator, where the gap is proportional to the condensate and hybridization parameter, and makes it amenable to a topological analysis. Since the electron states being hybridized have the opposite parities, the resulting Kondo insulator is topological and contains its hallmark feature - the Dirac surface states. Thus, the mean-field theory is reasonable, and works at the qualitative level. A fundamental quantity describing the magnetic response of the system is the spin susceptibility. The Kondo screening mechanisms are mostly characterized by the same response function. We calculate the bulk spin susceptibility of this system, when identified by the dominance of the bulk metallic character, in a sequel to this work.

The paper is organized as follows: In section 2, we consider a model for a (topological) Kondo insulator on a simple cubic lattice with one spin-degenerate orbital per lattice site each for *d* and *f* electrons with Hubbard type interaction term (*U*) between the latter. We also take into account the break-down of the time-reversal symmetry (TRS) induced by bulk substitutional magnetic impurities (MIs), such as Co atoms. We implement the slave-boson protocol to include the effect of infinite *U* into consideration at the mean-field theoretic (MFT) level. We obtain the grand canonical potential $\Omega$ of the system in the slave-boson representation with infinite *U*. The mean-field parameters, such as slave-boson field, auxiliary chemical potential, and a Lagrange multiplier to enforce the prohibition of double occupancy were obtained by minimizing this grand canonical potential with respect to these parameters. In section 3, we investigate the surface state dispersion followed by the plasmonics of the surface states. Considering the intra-band and inter-band transitions, we find possibility of collective mode exhibiting acoustic dispersion for the gapped surface states. This section also comprises of discussion on the confinement measure of the surface plasmons. The strong incarceration capability of these plasmons is a significant outcome of the present work. The paper ends with a short discussion and concluding remarks at the end.

## 2. PERIODIC ANDERSON MODEL

The widely accepted model of a topological Kondo insulator (TKI) **[7-10, 16]** involves, alongside a strong spin-orbit coupling, hybridization between an odd-parity nearly localized band and an even-parity delocalized conduction band. In the case of $SmB_6$ with a cubic crystal structure, these bands correspond to 4*f* and 5*d* electrons, respectively. It is then imperative that we start with the periodic Anderson model (PAM) where there are two different species of electrons, namely conduction electrons and localized electrons, often originating from *d* and *f* orbitals, respectively. We shall call them the conduction electrons and the valence electrons, respectively. The model **[1-6, 8, 25]** given below ignores the complicated multiplet structure of the *d* and *f* orbitals usually encountered in real TKIs such as $SmB_6$. This has no major implication as most topological properties of cubic Kondo insulators do not depend on the precise form of the hopping and hybridization matrix elements or, on the particular shape of the orbitals.

## 2.1 MODEL

We consider below a well-known model **[1-6,8]** for a (topological) Kondo insulator on a simple cubic lattice with one spin-degenerate orbital per lattice site each for *d* and *f* electrons. In momentum-space, we represent them by creation (annihilation) operators $d^\dagger_{k\zeta}$ ($d_{k\zeta}$) and $f^\dagger_{k\zeta}$ ($f_{k\zeta}$), respectively. Here, the index $\zeta$ (= ↑,↓) represents the spin or pseudo-spin of the electrons, where the latter is relevant for the localized electrons with generally large SOC. The Hamiltonian consists of two parts, namely, the bare hopping of the individual orbitals plus the hybridization between *d* and *f* orbitals(ℵ), and an onsite repulsion or *f* electrons(ℵ$_{int}$). We have

$$\aleph = \sum_{k\zeta\, =\uparrow,\downarrow}(-\mu - \epsilon^d_k)d^\dagger_{k,\zeta} d_{k,\zeta} + \sum_{k,\zeta\, =\uparrow,\downarrow}(-\mu - \epsilon^f_k) f^\dagger_{k,\zeta} f_{k,\zeta} + \sum_{k,\zeta=\uparrow,\downarrow}\{\Gamma_{\zeta\,=\uparrow,\downarrow}(k)\, d^\dagger_{k\zeta} f_{k,\zeta} + \text{H.C.}\} \quad (1)$$

Where $\epsilon^d_k = [2t_{d1} c_1(k) + 4t_{d2} c_2(k) + 8t_{d3} c_3(k)]$ and $\epsilon^f_k = [-\epsilon_f + 2t_{f1} c_1(k) + 4t_{f2} c_2(k) + 8t_{f3} c_3(k)]$. The $(t_{d1}, t_{f1})$, $(t_{d2}, t_{f2})$, and $(t_{d3}, t_{f3})$, respectively, are the *NN*, *NNN*, and *NNNN* hopping parameters. Also, $c_1(k) = (\cos k_x a + \cos k_y a + \cos k_z a)$, $c_2(k) = (\cos k_x a \cos k_y a + \cos k_y a \cos k_z a + \cos k_z a \cos k_x a)$, $c_3(k) = (\cos k_x a \cos k_y a \cos k_z a)$ with *a* as the lattice constant. The sums run over all values for the crystal momentum (*k*) and the index $\zeta = (\uparrow,\downarrow)$ for all (cubic) lattice sites. The dispersion of the *d* and *f* electrons is described by the first and the second term, respectively, and their hybridization or the spin-orbit interaction by the matrix $\Gamma_{\zeta=\uparrow,\downarrow}(k)$. Due to spin-orbit coupling the *f*-states are eigenstates of the total angular momentum *J*, and hence hybridize with conduction band states with the same symmetry. This gives rise to the momentum-dependence of the form-factor $\Gamma_{\zeta=\uparrow,\downarrow}(k)$. The spin–orbit interaction or hybridization term **[1, 21]** is given by $\Gamma_{\zeta=\uparrow,\downarrow}(k) = -2V(s(k)\cdot\zeta)$ where $\zeta_\alpha$ are the Pauli matrices in physical spin space, and $s(k) = (\sin k_x a, \sin k_y a, \sin k_z a)$. The hybridization is the source of non-trivial topology of the emergent bands. Since the *f*- and *d*-states have different parities, for the hybridization we must have odd parity: $\Gamma_{\zeta=\uparrow,\downarrow}(-k) = -\Gamma_{\zeta=\uparrow,\downarrow}(k)$. The negative sign of $t_{f1}$ is necessary for the band inversion, which induces the topological state **[27]**. The system shows the insulating as well as the metallic phases. The difference between the metallic and insulating phase is the sign of $t_{f1}$: It is positive for the metallic and negative for the insulating phase. The bandwidth of the *f* electrons is much smaller than the bandwidth of the conduction electrons and we therefore assume that $|t_{f1}| << |t_{d1}|$ and similar relations hold for second- and third-neighbor hopping amplitudes. The hybridization is characterized by the parameter *V* for which we typically use $|V| < |t_{d1}|$. Throughout the whole paper, we choose $t_{d1}$ to be the unit of energy, $t_{d1}= 1$. We note that the hybridization is an odd function of *k* in order to preserve time reversal symmetry(TRS), as it couples to the physical spin of the electron. The interaction term $\aleph_{int} = U \sum_i f^\dagger_{i\uparrow} f_{i\uparrow} f^\dagger_{i\downarrow} f_{i\downarrow}$ is the onsite repulsion of *f* electrons. Here we have assumed that the *f* electrons locally interact via a Hubbard-*U* repulsion while the *d* electrons are non-interacting. The total Hamiltonian (without the interaction term *U*) yields the single–particle spectrum $\epsilon^{(\zeta)}_\alpha(k) = -\frac{(\epsilon^d_k+\epsilon^f_k+\zeta M)}{2} + \alpha\sqrt{\frac{(\epsilon^d_k-\epsilon^f_k+\zeta M)^2}{4} + \epsilon^2_h}$ where $\alpha = 1\,(-1)$ for upper band (lower band) $\zeta = \pm 1$ labels the eigenstates (↑,↓) of $\zeta_z$, and $\epsilon_h = -2V(s_x^2 + s_y^2 + s_z^2)^{1/2}$. We refrain from showing the pictorial depiction of the spectrum. The interested reader may refer to the thesis of Legner [1].Our aim in this paper is to capture physics associated with the surface plasmon in the system, modelled by appropriate surface Hamiltonian derived from the bulk Hamiltonian $\aleph + \aleph_{int}$ above. The interaction term $\aleph_{int}$ can be studied non-perturbatively by various methods, including dynamical mean-field theory, Gutzwiller-projected variational wave-functions, or slave-particle representations [1, 29-32].In what follows we shall use the slave-particle protocol. The *f*-electron correlations, being larger than all other energy scales in the problem, effectively enforce a no-double-occupancy constraint on each site, and therefore contribute through virtual processes at low temperatures only. The original Hilbert space thus get projected onto a smaller subspace, where the double occupancy is excluded.

We now touch upon the break-down of TRS due to bulk substitutional MIs, such as Co atoms, for $t_{f1} > 0$. For $t_{f1} < 0$, the itinerant electrons may not be available. We model the interaction between an impurity moment and the itinerant (conduction) electrons in the system with coupling term $J \sum_m S_m \cdot s_m$, where $S_m$ is the *m* th-site impurity spin, $s_m = \left(\frac{1}{2}\right) d^\dagger_{m\zeta} \zeta_z d_{m\zeta}$, $d_{m\zeta}$ is the fermion annihilation operator at site-*m* and spin-state $\zeta$ (=↑,↓) and $\zeta_z$ is the z-component of the Pauli matrices. We make the approximation of treating the MI spins as classical vectors. The latter is valid for |***S***| >1. For a Co atom with the electronic structure as $4s^2 3d^7$, there are seven electrons in *d* orbitals with degenerate level of orbitals as five. There are three unpaired electrons with total spin angular momentum quantum number (3/2) and the total orbital quantum number (3). These values lead to total magnetic moment √27 Bohr magneton. Upon equating this with the formula for the spin quantum number only, the magnetic moment, viz. $\mu = \sqrt{\{4S(S+1)\}}$, we obtain $S \rightarrow 2.15, -3.15$ justifying the approximation made above. Absorbing the magnitude of the MI spin into the coupling constant $J$ ($M = J|S|/t_{d1}$)it follows that the exchange field term, in the $(d_{k\uparrow} f_{k\uparrow} d_{k\downarrow} f_{k\downarrow})^T$ basis, appears as $\{\zeta_z \otimes M(\tau_0 + \tau_z)/2\}$, where $\tau_0$ and $\tau_z$, respectively, are the identity and the z-component of Pauli matrix for the pseudo-spin orbital indices. We thus obtain the dimensionless contribution $[M \sum_{k,\zeta} sgn(\zeta) d^\dagger_{k,\zeta} d_{k,\zeta}]$ to the momentum space Hamiltonian above. We define the time reversal operator *T* as $T = i (\zeta_y \otimes \tau_0) \theta$. Here $U_T = i (\zeta_y \otimes \tau_0)$ is an unitary matrix and $\theta$ is the complex conjugation operator ($\theta = \theta^{-1}$) in the case of 4 ×4 matrices. This makes the time reversal anti-unitary. The time-reversal operator flips the sign of a spin ($T \zeta_j T^{-1} = -\zeta_j$ )where $j = (x,y,z)$, and also acting on ***k*** gives − ***k***. It follows that the exchange field term $\{\zeta_z \otimes M(\tau_0 + \tau_z)/2\}$ breaks TRS. The MIs do not act as scattering agent in our scheme; their effect is included in the band dispersion.

## 2.2 SLAVE-BOSON PROTOCOL

One utilizes the well-known slave-boson protocol [1,12-14] to do the projection onto the Hilbert space. In this protocol, the electron operator is expressed in terms of psuedo-fermions and slave-bosons. The operators $a^\dagger$ (heavy slave-boson creation operator) and $b^\dagger$ (light slave-boson creation operator), respectively, creates doubly-occupied and empty bosonic impurity states out of vacuum. The bosons created by $b^\dagger$ are supposed to carry the electron's charge. The single site fermionic occupation operator is denoted by $s^\dagger$. This corresponds to a fermionic operator and carry the electron's spin but no charge. For $U \rightarrow \infty$, the double occupancy is prohibited, and hence the operators $a^\dagger$ and $a$ will not be under consideration. The connection of the remaining auxiliary operators ( *b, s*) to the physical *f*- electron operator is $f^\dagger_\zeta = s^\dagger_\zeta b$ .In the slave-boson mean-field theory of the infinite-*U* model ( which consists of replacing slave-boson field (*b*) at each lattice site by the modulus of its expecetation value – a *c* number), the anti-commutation relation $\{ f_\zeta, f^\dagger_{\zeta'} \} = \delta_{\zeta\zeta'}$, implies that $b^2 \{ s_\zeta, s^\dagger_{\zeta'} \} = \delta_{\zeta\zeta'}$. To take care of the conservation of auxiliary particle number one needs to impose the restriction $\sum_\zeta s_\zeta^\dagger s_\zeta + b^\dagger b = 1$ or, $\sum_\zeta \langle s_\zeta^\dagger s_\zeta \rangle \cong 1 - b^2$ at a site. It is now straight-forward to write the thermal average of the TKI slave-boson mean-field Hamiltonian

$$\langle \aleph_{sb}(b,\lambda,\xi) \rangle = \sum_{k\zeta=\uparrow,\downarrow}(-\mu - \xi - \epsilon_k^d)\langle d_{k,\zeta}^\dagger d_{k,\zeta} \rangle + \sum_{k,\zeta=\uparrow,\downarrow}(-\mu + \xi - b^2\epsilon_k^f + \lambda)\langle s_{k,\zeta}^\dagger s_{k,\zeta} \rangle$$

$$+ b\sum_{k,\zeta,\sigma=\uparrow,\downarrow}\{\Gamma_{\zeta=\uparrow,\downarrow}(\mathbf{k})\langle d_{k\zeta}^\dagger s_{k,\zeta} \rangle + \text{H.C.}\} + \lambda N_s(b^2 - 1) \quad (3)$$

where the additional terms, in comparison with (1), are $-\mu((N_d + N_s) - N) - \xi(N_d - N_s) + \lambda[\sum_{k,\sigma=\uparrow,\downarrow}\langle s_{k,\zeta}^\dagger s_{k,\zeta} \rangle + N_s(b^2 - 1)]$. The first term is the constraint which fixes the total number of particles $N$ ($N = N_d + N_s$), the second term enforces the fact that there are equal $d$ and $s$ fermions ($N_d = N_s$), and the third term describes the constraint on the pseudo-particles due to the infinite Coulomb repulsion. In order to have a Kondo insulator, formation of singlet states between $d$ and $s$ fermions is needed at each lattice site. This means that the number of $d$ and $s$ fermions are equal on average. Here $\lambda$ is a Lagrange multiplier, and $N_s$ is the number of lattice sites for $s$ electrons (similarly, $N_d$ corresponds to $d$-electrons). The dispersion of the $f$-electron is renormalized by $\lambda$ and its hopping amplitude by $b^2$. Moreover, the hybridization amplitude is also renormalized by the $c$-number $b$. The total parameters are slave-boson field $b$, auxiliary chemical potentials $\xi$ and $\mu$ ($\mu$ is a free parameter), and the Lagrange multiplier $\lambda$. Though these facts are explained clearly in ref. **[1]**, we have never-the-less found necessary to include them to make this paper self-contained. One obtains equations for the parameters ($b, \lambda, \xi$) minimizing the thermodynamic potential per unit volume $\Omega_{sb} = -(\beta V)^{-1} \ln Tr \exp(-\beta\aleph_{sb}(b,\lambda,\xi))$ : $\partial\Omega_{sb}/\partial b = 0$, $\partial\Omega_{sb}/\partial\lambda = 0$, and $\partial\Omega_{sb}/\partial\xi = 0$. Here $\beta$ denotes the inverse of the product of temperature $T$ and Boltzmann constant $k_B$. These equations are (see Appendix A **[23,24]** for the calculation of $\Omega_{sb}$) :

$$2\lambda b = N_s^{-1} \sum_k \frac{\partial}{\partial b}[2V(s_x - is_y)\langle d_{k\uparrow}^\dagger bs_{k,\downarrow}\rangle + 2V(s_x - is_y)\langle bs^\dagger_{k,\uparrow} d_{k\downarrow}\rangle + \text{H.C.}]$$

$$+ N_s^{-1}\sum_{k,\zeta}\frac{\partial}{\partial b}[\epsilon_k^f \langle bs_{k,\zeta}^\dagger bs_{k,\zeta}\rangle + \epsilon_k^d \langle d_{k,\zeta}^\dagger d_{k,\zeta}\rangle] + N_s^{-1}\sum_{k,\zeta}\frac{\partial}{\partial b}[2Vs_z\zeta \langle d_{k,\zeta}^\dagger bs_{k,\zeta}\rangle + \text{H.C.}], \quad (4)$$

$$(1 - b^2) = N_s^{-1}\sum_k \frac{\partial}{\partial\lambda}[2V(s_x - is_y)\langle d_{k\uparrow}^\dagger bs_{k,\downarrow}\rangle + 2V(s_x - is_y)\langle bs^\dagger_{k,\uparrow} d_{k\downarrow}\rangle + \text{H.C.}]$$

$$+ N_s^{-1}\sum_{k,\zeta}\frac{\partial}{\partial\lambda}[\epsilon_k^f \langle bs_{k,\zeta}^\dagger bs_{k,\zeta}\rangle + \epsilon_k^d\langle d_{k,\zeta}^\dagger d_{k,\zeta}\rangle] + N_s^{-1}\sum_{k,\zeta}\frac{\partial}{\partial\lambda}[2Vs_z\zeta\langle d_{k,\zeta}^\dagger bs_{k,\zeta}\rangle + \text{H.C.}], \quad (5)$$

$$0 = N_s^{-1}\sum_k \frac{\partial}{\partial\xi}[2V(s_x - is_y)\langle d_{k\uparrow}^\dagger bs_{k,\downarrow}\rangle + 2V(s_x - is_y)\langle bs^\dagger_{k,\uparrow} d_{k\downarrow}\rangle + \text{H.C.}]$$

$$+ N_s^{-1}\sum_{k,\zeta}\frac{\partial}{\partial\xi}[\epsilon_k^f\langle bs_{k,\zeta}^\dagger bs_{k,\zeta}\rangle + \epsilon_k^d\langle d_{k,\zeta}^\dagger d_{k,\zeta}\rangle] + N_s^{-1}\sum_{k,\zeta}\frac{\partial}{\partial\xi}[2Vs_z\zeta\langle d_{k,\zeta}^\dagger bs_{k,\zeta}\rangle + \text{H.C.}]. \quad (6)$$

$$\epsilon_k^d = [2t_{d1}c_1(k) + 4t_{d2}c_2(k) + 8t_{d3}c_3(k)], \quad (7)$$

$$\epsilon_k^f = [-\epsilon_f + 2t_{f1}c_1(k) + 4t_{f2}c_2(k) + 8t_{f3}c_3(k)], \quad (8)$$

The averages $\langle d^\dagger_{k,\zeta} d_{k,\zeta}\rangle$, $\langle bs^\dagger_{k,\zeta} bs_{k,\zeta}\rangle$, etc. have been calculated in appendix B below. Their expressions show that in the zero-temperature and the long-wavelength limits, the contribution of the averages $\langle d^\dagger_{k\uparrow} bs_{k,\downarrow}\rangle$ and $\langle bs^\dagger_{k,\uparrow} d_{k\downarrow}\rangle$, etc. to the derivatives in Eqs. (4), (5), and (6) are insignificant in comparison with those of $\sum_{k,\zeta}[\epsilon^f_k \langle bs^\dagger_{k,\zeta} bs_{k,\zeta}\rangle + \epsilon^d_k \langle d^\dagger_{k,\zeta} d_{k,\zeta}\rangle]$. This observation allows us to approximate the equations as

$$2\lambda b \approx N_s^{-1} \sum_{k,\zeta} \frac{\partial}{\partial b}[\epsilon^f_k \langle bs^\dagger_{k,\zeta} bs_{k,\zeta}\rangle + \epsilon^d_k \langle d^\dagger_{k,\zeta} d_{k,\zeta}\rangle], \tag{9}$$

$$(1-b^2) \approx N_s^{-1} \sum_{k,\zeta} \frac{\partial}{\partial \lambda}[\epsilon^f_k \langle bs^\dagger_{k,\zeta} bs_{k,\zeta}\rangle + \epsilon^d_k \langle d^\dagger_{k,\zeta} d_{k,\zeta}\rangle], \tag{10}$$

$$0 \approx N_s^{-1} \sum_{k,\zeta} \frac{\partial}{\partial \xi}[\epsilon^f_k \langle bs^\dagger_{k,\zeta} bs_{k,\zeta}\rangle + \epsilon^d_k \langle d^\dagger_{k,\zeta} d_{k,\zeta}\rangle] \tag{11}$$

in these limits. As we already have noted, for the conservation of auxiliary particle number one needs to impose the restriction $\int d\mathbf{r} \sum_\zeta \langle bs^\dagger_\zeta(\mathbf{r}) bs_\zeta(\mathbf{r})\rangle \cong N_s b^2(1-b^2)$. Here $s_\zeta(\mathbf{r}) = N_s^{-\frac{1}{2}} \sum_k e^{i\mathbf{k}\cdot\mathbf{r}} s_{k,\zeta}$. This is the fourth equation with (9)-(11) as the first three and we have four unknowns, viz. ($b$, $\lambda$, $\xi$, $ak_F$) where $a \approx 4.13$ Å is the lattice constant and $k_F$ is the Fermi wave number. In view of the fact that the chemical potential is a free parameter and could be somewhere between the valence and the conduction bands ($\mu > \epsilon^{(+)}_-(k), \epsilon^{(-)}_-(k)$), once again it is easy to see that for the low-lying states in the zero-temperature limit one may write the imposed restriction as $2 N_s^{-1} \sum_{k,\zeta} 1 = b^2 - b^4$. This is an equation for $b^2$ in terms of $(ak_F)$. After a little algebra, we find that whereas (9) and (10) together yield $\lambda = -6t_{f1} + 6b^2 t_{f1}$, Eq.(11) yields $\xi = -3t_{d1} + 3t_{f1} + M^2/\{4(\mu + 3t_{d1} + 3t_{f1})\}$ where the exchange field $M$ is also a free parameter. We estimate $(ak_F)$ in the following manner: Since the Fermi velocity $v_F^*$ of the low-lying states is known to be less than 0.2 eV-Å[25], taking the effective fermion mass ($m^* = \frac{\hbar k_F}{v_F^*}$) a hundred times that of an electron[26], we find that $(ak_F) \sim 0.01$. This is consistent with the long wavelength limit we have assumed. The two values of $b^2$ obtained from here are close to but less than one ($1^-$) and close to zero ($0^+$). It may be noted now that when $b$ is nonzero, the system is in Kondo state, and when $b$ vanishes, the system is in normal gas state. The admissible value of $b^2$ will be thus be $1^-$. With this a reasonable spectral gap, in the long wavelength limit, is obtained in the Kondo insulating state. We have seen that in order to have a gapped Kondo state it is required that the tunneling of the localized fermions must have an opposite sign compared with the conduction fermions, which is difficult to achieve, because the signs of the tunneling for the lowest Bloch bands are usually the same. However, here it will be taken to be positive for the metallic phase and negative for the insulating phase, as for the former phase the leading term for f-electrons in the Hamiltonian, viz. $-2t_{f1} c_1(k) \langle bs^\dagger_{k,\zeta} bs_{k,\zeta}\rangle$, corresponds to a minimum at (0,0,0) (electron-like band) and in the latter phase corresponds to a maximum (hole-like band). It must be mentioned that in their quantum simulation of the topological Kondo insulator in ultra-cold atoms, Zheng et al [27] have suggested a transformation leading to a staggered Kondo coupling – positive for the metallic phase and negative for the insulating phase. As regards $\xi$, it is $[-2.7 + M^2/\{4(\mu + 3.3)\}]$ in the metallic phase and $[-3.3 + M^2/\{4(\mu + 2.7)\}]$ in the insulating phase. As all the unknown parameters have been determined now, the stage is set to investigate the surface state plasmon. Before we do this an important point to be noted here is that the thermodynamic potential per unit volume for $\Omega_{sb}$ for the metallic phase is less than that at the insulating phase which indicates that the ground state of TKI bulk is metallic in the presence of strong electron correlation and the metallicity survives even if the onsite repulsion of f electrons is much larger compared to $t_{d1}$.

## 3. SURFACE STATES AND PLASMONS

### 3.1 Surface states

In order to set the stage, we choose a representation involving the states ($|c_{k,\uparrow,\pm}\rangle, |c_{k,\downarrow,\pm}\rangle$), where the operators $c_{k,\uparrow,\pm}$ and $c_{k,\downarrow,\pm}$, respectively, are the spin-up and spin-down annihilation operators corresponding to the upper and lower bands with spin-splitting ($\epsilon^{(\uparrow)}_{\pm}(k)$ and $\epsilon^{(\downarrow)}_{\pm}(k)$), as our basis. In the basis chosen above our bulk mean-field Hamiltonian matrix in slave-boson protocol will appear as

$$\begin{pmatrix} -\mu - \xi - \varepsilon_k^d + M & -2Vbs_z & 0 & -2Vb(s_x - is_y) \\ -2Vbs_z & -\mu + \xi - b^2\varepsilon_k^f + \lambda & -2Vb(s_x - is_y) & 0 \\ 0 & -2Vb(s_x + is_y) & -\mu - \xi - \varepsilon_k^d - M & 2Vbs_z \\ -2Vb(s_x + is_y) & 0 & 2Vbs_z & -\mu + \xi - b^2\varepsilon_k^f + \lambda \end{pmatrix} \quad (12)$$

The eigenvalues of this matrix are given by (B9). The energy eigen vectors corresponding to these eigenvalues $\left(\epsilon_\alpha^{(\zeta)}(k)\right)$ are given by

$$\Psi_{k\alpha}^{(\zeta=-1)} = N_1^{-1/2} \begin{pmatrix} E_\alpha^{(\zeta=-1)}(k) \\ 2Vbs_z \\ 0 \\ 2Vb(s_x + is_y) \end{pmatrix},$$

$$\psi_{k\alpha}^{(\zeta=+1)} = N_2^{-1/2} \begin{pmatrix} 0 \\ -2Vb(s_x - is_y) \\ -E_\alpha^{(\zeta=1)}(k) \\ 2Vbs_z \end{pmatrix}. \quad (13)$$

where

$(N_1, N_2) = \left[E_\alpha^{(\zeta=\mp 1)^2}(k) + 4V^2 b^2(s^2_x + s^2_y + s^2_z)\right]^{\frac{1}{2}}$ correspond to the normalization terms and $E_\alpha^{(\zeta)}(k) = -\left(\varepsilon_f + \epsilon_\alpha^{(\zeta)}(k)\right) = \xi - b^2\epsilon_k^f + \lambda - \epsilon_\alpha^{(\zeta)}(k)$. An important physical consequence of the non-trivial topological band structure is the existence of Dirac-like surface states with helical spin texture. Our aim in this section is to examine the surface plasmons of the system under consideration. The plasmons are defined as longitudinal in-phase oscillation of all the carriers driven by the self-consistent electric field generated by the local variation in charge density. We require surface state single particle excitation spectrum for this purpose. To study the surface states, we first consider a thick slab limited in $z \in [-d/2, d/2]$ with open boundary conditions, where $d$ is the thickness of the slab in $z$ direction. In this case $k_z$ is not a good quantum number which should be replaced by $-i\partial_z$. The Hamiltonian $\aleph^{slab}(k_x, k_y)$ for the slab structure under consideration can be obtained from the Hamiltonian $\aleph^{bulk}(k_x, k_y, -i\partial_z)$ considering the ortho-normal function $|\varphi_n(z)\rangle = \psi_n(z)$ where $\psi_n(z) = \left(\frac{2}{d}\right)^{\frac{1}{2}}\left[\sin\left\{\frac{n\pi\left(z+\frac{d}{2}\right)}{d}\right\}\right]$ $(n = 0,1,2,3,\ldots)$ ensuring $\psi_n\left(z = -\frac{d}{2}, \frac{d}{2}\right) = 0$. The matrix element may be written as

$$\aleph_{mn}{}^{slab}(k_x, k_y) = \int_{-d/2}^{d/2} \langle \varphi_m(z) \mid \aleph_{mn}{}^{bulk}(k_x, k_y, -i\partial_z) \rangle \mid \varphi_n(z)\rangle dz. \tag{14}$$

The eigenstates of $\aleph_{mn}{}^{slab}(k_x, k_y)$ are the so called edge states. Equation (12) yields $\aleph_{mn}{}^{slab}$ as

$$\begin{pmatrix} \Gamma_1 & -\frac{16iVba}{3d} & 0 & 0 \\ \frac{16iVba}{3d} & \Gamma_2 & 0 & 0 \\ 0 & 0 & \Gamma_3 & \frac{96iVba}{7d} \\ 0 & 0 & -\frac{96iVba}{7d} & \Gamma_2 \end{pmatrix} \tag{15}$$

where

$$\Gamma_1(K) = -\mu - \xi - 6t_{d1} + t_{d1}K^2 + M,$$

$$\Gamma_2(K) = -\mu + \xi + b^2\varepsilon_f - 6b^2 t_{f1} + b^2 t_{f1}K^2 + \lambda,$$

$$\Gamma_3(K) = -\mu - \xi - 6t_{d1} + t_{d1}K^2 - M. \tag{16}$$

Upon noting that $\lambda = -6t_{f1} + 6b^2 t_{f1}$ and $\xi \approx -3t_{d1}$, the eigenvalues of this matrix is given by

$$\epsilon_{\alpha,\,surface}^{(\zeta)}(K) = -\mu + \left(-3t_{d1} - 3t_{f1} + \frac{b^2}{2}\varepsilon_f + \frac{\zeta M}{2}\right) + \frac{t_{d1}+b^2 t_{f1}}{2}(Ka)^2 + \alpha$$

$$\times \sqrt{(CVb)^2 + \frac{b^4}{4}\varepsilon_f^2 + \left(3t_{f1} + \frac{\zeta M}{2}\right)^2 - b^2\varepsilon_f\left(3t_{f1} + \frac{\zeta M}{2}\right) + v_F^*(\zeta)^2(Ka)^2 + O((Ka)^4)}$$

where $v_F^*(\zeta) = \sqrt{(t_{d1} - b^2 t_{f1})\left(3t_{f1} + \frac{\zeta M}{2} - \frac{b^2}{2}\varepsilon_f\right)}$ will be interpreted as the Fermi velocity. The constant $C = \left(\frac{16a}{3d}\right)$ for $\zeta = 1$ and $\left(\frac{96a}{7d}\right)$ for $\zeta = -1$. Thus, the intervention of the magnetic impurities on the surface imparts different velocities to the up and down quasi-particle spins. The eigenvectors corresponding to the eigenvalue $\epsilon_{\alpha,\,surface}^{(\zeta=+1)}(K)$ are

$$\tilde{V}_{k\alpha}^{(\zeta=+1)} = N_1^{-1/2}\begin{pmatrix} -\xi - b^2\varepsilon_f + 6b^2 t_{f1} - b^2 t_{f1}K^2 - \lambda + \epsilon_{\alpha,surface}^{(\zeta=+1)}(K) \\ 16Vbai/3d \\ 0 \\ 0 \end{pmatrix}, \tag{17}$$

and those corresponding to the eigenvalue $\epsilon_{\alpha,\,surface}^{(\zeta=-1)}(K)$ are

$$\tilde{V}_{k\alpha}^{(\zeta=-1)}= N_2^{-1/2} \begin{pmatrix} 0 \\ 0 \\ \xi + b^2\varepsilon_f - 6b^2 t_{f1} + b^2 t_{f1} K^2 + \lambda - \in_{\alpha,surface}^{(\zeta=-1)}(K) \\ 96Vbai/7d \end{pmatrix}. \quad (18)$$

where $(N_1, N_2)$ correspond to the normalization terms. Taking $M = 0$, we notice that if the $f$-electrons somehow satisfy the condition $(3t_{f1} - \frac{b^2}{2}\varepsilon_f + \frac{(CVb)^2}{3t_{f1}-\frac{b^2}{2}\varepsilon_f})$ small compared to $(t_{d1} - b^2 t_{f1}) \times (aK)^2$, one obtains gapless surface state bands at (0,0) wavevector

$$\in_{\alpha,\ surface}(K) \approx [-\mu + (-3t_{d1} - 3t_{f1} + \frac{b^2}{2}\varepsilon_f) + \alpha v_F^* \mid aK \mid + \frac{t_{d1}+b^2 t_{f1}}{2}(Ka)^2] \quad (19)$$

for the insulating bulk. For this to happen, the severe restrictions are

$$\mid \varepsilon_f \mid > \frac{6|t_{f1}|}{b^2} \approx 6.5|t_{f1}|. \quad (20)$$

with $b^2 = 0.9890$ (i.e. $\mid t_{f1} \mid$ should be less than $\mid \varepsilon_f \mid$ by one order of magnitude) and the high symmetry point $(0,0)$ is unapproachable as

$$\mid aK \mid > \sqrt{\frac{3t_{f1}+v-\frac{b^2}{2}\varepsilon_f}{(t_{d1}-b^2 t_{f1})}}. \quad (21)$$

Here $v = \frac{(\frac{96a}{7d}Vb)^2}{3t_{f1}-\frac{b^2}{2}\varepsilon_f}$. The Fermi velocity $v_F^*$ is given by $\sqrt{(t_{d1} + b^2 \mid t_{f1} \mid)(-3 \mid t_{f1} \mid + \frac{b^2}{2} \mid \varepsilon_f \mid)}$. With the parameter choice $ak_F = 0.01, t_{d1} = 500\ meV$, $\mid t_{f1} \mid = 5\text{meV}$, $t_{d2} = t_{d3} = t_{f2} = t_{f3} = 0$, $V = 100\ meV, b^2 = 0.9890$, and $\mid \varepsilon_f \mid = 50$ meV, we find the estimated value $v_F^* \approx 10^4 m - s^{-1}$. The Kondo screening length $(= \frac{\hbar v_F^*}{k_B T_{K,s}})$ for the surface states will be, therefore, be one order of magnitude higher than the lattice constant($a$) for the surface Kondo temperature $T_{K,s} \sim 20K$. Note that the surface Kondo temperature is lower than the bulk Kondo temperature due to the reduction in the Kondo coupling constant. This happens due to the reduction in the $f$-electron coordination number and the internal gapping of the bulk states.

There is continuing row **[28,29]** over whether the origin of observed anomalies at low temperatures in SmB$_6$, such as finite linear specific heat coefficient, bulk optical conductivity below the charge gap, etc., correspond to the bulk metallicity or the surface metallicity with the insulating bulk. As we have now access to the bulk dispersion given by (B9), and the surface metallicity given by Eq. (19) albeit under a condition demanding strong $f$-electron localization, these issues could be looked into. We, however, consider the case when the restrictions given by (20) and (21) are not satisfied. For $M = 0$, we then have

$$\in_{\alpha,\zeta,surface}(K) = -\mu + (-3t_{d1} - 3t_{f1} + \frac{b^2}{2}\varepsilon_f) + \frac{t_{d1}+b^2 t_{f1}}{2}(Ka)^2$$

$$+\alpha \sqrt{(CVb)^2 + (\frac{b^2}{2}\varepsilon_f - 3t_{f1})^2 + v_F^*(\zeta)^2 (Ka)^2} + O((Ka)^4), \quad (22)$$

We investigate the possibility of the surface plasmons with the gapped excitation spectrum given by Eq. (22) as, unlike Eq.(19) which is expected to give conventional $q^{1/2}$ plasmons, this dispersion is expected to yield plasmons with unconventional dispersion relation. The detection of such plasmons experimentally establishes the dominance of the bulk metallicity. Furthermore, it must be noted that Legner et al. [26] have found an expression of the Fermi velocity which with the same set of parameter values as above yields $v_F^*$ nearly the same value as obtained by us. The expression $v_F^* = \sqrt{(t_{d1} - b^2 t_{f1})(3t_{f1} - \frac{b^2}{2}\varepsilon_f)}$ indicates that, apart from the hopping for the Bloch bands, the existence of the conducting surface for the insulating bulk is rooted in the fundamental conditions of the strong electron correlation and the localization of *f*-electrons($\varepsilon_f < 0$) for a Kondo system. The hybridization parameter *V* cannot contribute here as we have chosen an ortho − normal function to obtain the Hamiltonian for the slab structure from the bulk Hamiltonian. The discussion above allows us to infer that a bulk TKI evolves from a metal to an insulator with a small gap as the sign of $t_{f1}$ is changed from positive to negative. It could be a topological insulator with a metallic surface state provided $|t_{f1}| \ll |\varepsilon_f|$. Since the calculation above is in the zero –temperature limit, a prediction of the previous works[7-10] that "when temperature is lowered a Kondo insulator may turn into a topological insulator with a metallic surface state" remains unverified. However, our finding of surface states with gapless Dirac dispersion corroborates an important experimental finding of Xiang et al.[28]. They have found that, in the case of the prototypical TKI $SmB_6$, there is a broken rotating symmetry in the amplitude of the main de Haas–van Alphen oscillation branch consistent with Lifshitz-Kosevich theory confirming a 2D nature of the electronic state, similar to quantum oscillation experiments [29,30] for the conventional TIs to probe the surface Dirac fermions. The transport measurements [31,32] in the past have also demonstrated the insulating bulk and metallic surface separation. For potential applications toward scalable quantum information processing [33] this bulk and surface separation is especially important. One pertinent question is can one add an extra term involving *M* to open a gap at the surface states with gapless Dirac dispersion. From the surface energy eigenvalues above, it is clear that the answer is 'no'('yes') as long as we intend to preserve (break) the time reversal symmetry(TRS). In other words, if the system preserves (breaks) the TRS, we have topological (trivial) insulator with protected (unprotected) edge states. In the latter case, since we are able to open a gap for the edge states, we can move the chemical potential into the gap and then the system turns into a trivial insulator.

### 3.2 Plasmon frequency

The plasmons are defined as longitudinal in-phase oscillation of all the carriers driven by the self-consistent electric field generated by the local variation in induced charge density $\rho(r,\omega)$. In a linear-response approximation, we have $\rho(\boldsymbol{r},\omega) = e^2 \int d^2\boldsymbol{r}'\chi(\boldsymbol{r},\boldsymbol{r}',\omega)\Phi(\boldsymbol{r}',\omega)$ where $\Phi$ is induced local potential and $\chi$ is the fermion response function or the dynamic polarization. This is a quantity of interest for many physical properties, since it determines e.g. the plasmon and phonon spectra. Assuming plasmon oscillation for the 3D system under consideration entirely a surface phenomenon, in the random phase approximation (RPA) [34,35], we write the dynamical polarization function $\chi(a\boldsymbol{q},\omega)$ in the momentum space, as

$$\chi(a\boldsymbol{q},\omega) = \sum_{K,\zeta,\zeta',\alpha,\alpha'} |\langle \Psi_{\zeta,\alpha}(a(\boldsymbol{K}-\boldsymbol{q}))| \Psi_{\zeta',\alpha'}(a\boldsymbol{K})\rangle|^2 \left[\frac{n_{\zeta,\alpha}(a\boldsymbol{K}-a\boldsymbol{q}) - n_{\zeta',\alpha'}(a\boldsymbol{K})}{\{\hbar\omega + \epsilon_{\zeta,\alpha,surface}(a\boldsymbol{K}-a\boldsymbol{q}) - \epsilon_{\zeta',\alpha',surface}(a\boldsymbol{K}) + i\eta\}}\right].$$

(23)

The symbol $\epsilon_{\zeta,\alpha,surface}(a\boldsymbol{K})$ stands for the surface state single-particle excitation spectrum given by (22) and $|\langle \Psi_{\zeta,\alpha}(a(\boldsymbol{K}-\boldsymbol{q}))| \Psi_{\zeta',\alpha'}(a\boldsymbol{K})\rangle|^2$ for the band-overlap of wave functions. Since we are presently interested on intra-band plasmons only, we write down the explicit expression for the intra-band overlap. In view of (17) and (18), this is given by $F_{\alpha,\alpha,\zeta,\zeta'}(K,q) = \left(\frac{1}{2}\right)[1 + \zeta\zeta' \cos\theta_{\alpha,\alpha,K,q}]$, where

$$\cos \theta_{\alpha,\alpha,K,q} = \left[\frac{(aq)\left(\frac{\partial \epsilon_{\alpha,surface}(K)}{\partial K} - \frac{\partial \Gamma_2(K)}{\partial K}\right)\left(\epsilon_{\alpha,surface}(K) - \Gamma_2(K)\right)}{\left\{\left(\epsilon_{\alpha,surface}(K) - \Gamma_2(K)\right)^2 + (CVb)^2\right\}}\right]. \tag{24}$$

in the long wavelength limit. Obviously enough, for the spin-split conduction / valence band, $\zeta\zeta' = -1$ and in the intra-band case it is +1. The occupation function for the band $\alpha = \pm 1$ is given by $n_{\zeta,\alpha}(aK) = [\exp(\beta\,(\epsilon_{\zeta,\alpha,surface}(aK)) - \beta\mu) + 1]^{\wedge} - 1$. The real-part of the polarization function

$$\chi_1(aq,\omega) \text{ is equal to } P\sum_{K,\alpha,\alpha',\zeta,\zeta'}\left[\frac{(n_{\zeta,\alpha}(aK-aq) - n_{\zeta',\alpha'}(aK))F_{\alpha,\alpha',\zeta,\zeta'}(K,q)}{\{\hbar\omega + \epsilon_{\zeta,\alpha,surface}(aK-aq) - \epsilon_{\zeta',\alpha',surface}(aK)\}}\right].$$

The imaginary part is given by

$$\chi_2(a\delta q,\omega') = -\pi \times$$
$$\sum_{K,\alpha,\alpha',\zeta,\zeta'}(n_{\zeta,\alpha}(aK-aq) - n_{\zeta',\alpha'}(aK))F_{\alpha,\alpha',\zeta,\zeta'}(K,q)\delta(\hbar\omega + \epsilon_{\zeta,\alpha,surf}(aK-aq) - \epsilon_{\zeta',\alpha',surf}(aK)).$$
$$\tag{25}$$

Here we have used the Sokhotski-Plemelj identity $(x \pm i\eta)^{-1} = P(x^{-1}) \mp i\pi\delta(x)$ with $P$ as the principal part. Since we have chosen $t_{d1}$ to be the unit of energy before, the quantity $\hbar\omega$ is also in the same unit. In the long-wavelength limit, the band structure in Eq.(22) yields

$$\epsilon_{\zeta,\alpha,surface}(aK-aq) - \epsilon_{\zeta,\alpha,surface}(aK) \approx \left\{\frac{a^2(q^2 - 2q.K)}{2\lambda_{\zeta,\alpha}}\right\}.$$

where

$$\lambda_{\zeta,\alpha} \approx \frac{\frac{2}{t_{d1}}}{\left\{1 + \frac{b^2 t_{f1}}{t_{d1}} + \frac{\alpha\,v_F^*(\zeta)^2}{2t_{d1}\left[(CVb)^2 + \left(\frac{b^2}{2}\varepsilon_f - 3t_{f1}\right)^2\right]^{\frac{1}{2}}}\right\}^{\frac{1}{2}}}.$$

Since throughout the paper we have chosen $t_{d1} = 1$ to be the unit of energy, therefore, $\lambda_{\zeta,\alpha}$ in this unit could be further approximated as

$$\lambda_{\zeta,\alpha} \approx 2(1 - b^2 t_{f1})\left\{1 - \frac{\left(-3|t_{f1}| + \frac{b^2}{2}|\varepsilon_f|\right)}{2\left[(CVb)^2 + \left(-3|t_{f1}| + \frac{b^2}{2}|\varepsilon_f|\right)^2\right]^{\frac{1}{2}}} + O\left((b^2 t_{f1})^2\right)\right\} \tag{26}$$

for the conduction electrons.

Within the random phase approximation (RPA), the plasmon dispersion is obtained by finding zeros of the dynamical dielectric function, which is expressed in terms of Cou-lomb's potential as $e_{\zeta,\alpha}(a|q|,\omega') = 1 - V(q)\chi_{\zeta,\alpha}(a|q|,\omega')$ where $\omega' = \omega - i\gamma$, $\gamma$ is the decay rate of plasmons, the expression $V(q) = \left(\frac{e^2}{2\varepsilon_0\varepsilon_r t_{d1} a^2 q}\right)$ is the Fourier transform of the Coulomb potential in two dimensions in the units of energy chosen to be $t_{d1} = 1$, $\varepsilon_0$ is the vacuum permittivity, and $\varepsilon_r$ is the relative permittivity of the surrounding medium. We have divided by surface area $a^2$ above to make $V(q)$ dimensionless. For weak damping, the equation $Re\,e(\omega,a|q|) = 0$ yields the plasmon frequency $\hbar\omega_{pl}$. In the intra-band case, from the equation above, in the high frequency limit we have

$$e_{\zeta,\alpha}(a|\boldsymbol{q}|,\omega) = 1 - V_0 \sum_{K,\alpha,\zeta} \left[ \frac{(\hbar\omega)^{-1} \frac{\partial n_{\zeta,\alpha}}{\partial \mu} \frac{\partial \epsilon_{\zeta,\alpha,surface}(aK)}{\partial(aK)} F_{\alpha,\zeta}(K,q)}{\left\{1 - \left\{\frac{(aq.aK)}{\hbar\omega\,\lambda_{\zeta,\alpha}}\right\}\right\}} \right]. \quad (27)$$

where $V_0 = \left(\frac{e^2}{2\varepsilon_0 \varepsilon_r a^2}\right)$ in the unit of energy, viz.. $t_{d1}$. The denominator of the summand in (27) involves a scalar product. We make use of the standard integral

$$\int_0^{2\pi} \frac{d\varphi}{\{a - x\cos\varphi\}} = 2\pi(2\theta(a)-1)\left(|a|^2 - x^2\right)^{-1/2}, \text{ for } |a| > x,$$

and zero, for $|a| \leq x$. Here $\theta(a)$ is the Heaviside step function. We can write $\int_0^{2\pi} \frac{d\varphi}{\left\{a - \left\{\frac{a|q||a||k|}{\hbar\omega\,\lambda_{\zeta,\alpha}}\right\}\cos\varphi\right\}}$

$=2\pi\left\{1 - \left(\frac{a|q||a||k|}{\hbar\omega\,\lambda_{\zeta,\alpha}}\right)^2\right\}^{-1/2}$ This integral allow us to write Eq.(27), for the conduction electrons, in the high frequency limit as

$$e_{\zeta,\alpha}(a|\boldsymbol{q}|,\omega) = 1 - 2\pi V_0 \sum_{K,\alpha,\zeta} \left[ \frac{(\hbar\omega)^{-1} \frac{\partial n_{\zeta,\alpha}}{\partial \mu} \frac{\partial \epsilon_{\zeta,\alpha,surface}(aK)}{\partial(aK)} F_{\alpha,\zeta}(K,q)}{\left\{1 - \left(\frac{a|q||a||k|}{\hbar\omega\,\lambda_{\zeta,\alpha}}\right)^2\right\}^{1/2}} \right]. \quad (28)$$

The denominator of the summand in (28) can be expanded using the Binomial theorem. We obtain then the following implicit equation as the Plasmon dispersion for the system:

$$(\hbar\omega)^3 - (\hbar\omega)^2 \left\{\pi V_0 \int d\boldsymbol{K} \sum_{\alpha,\zeta} \frac{\partial n_{\zeta,\alpha}}{\partial \mu} \frac{\partial \epsilon_{\zeta,\alpha,surface}(aK)}{\partial(aK)} (1 + (aq)\mathbb{F}_K)\right\}$$

$$= \frac{1}{2}\pi V_0 (aq)^2 \int d\boldsymbol{K} \sum_{\alpha,\zeta} \frac{\partial n_{\zeta,\alpha}}{\partial \mu} \frac{\partial \epsilon_{\zeta,\alpha,surface}(aK)}{\partial(aK)} \left(\frac{a|K|}{\lambda_{\zeta,\alpha}}\right)^2 (1 + (aq)\mathbb{F}_K), \quad (29)$$

where $d\boldsymbol{K} = \left(\frac{d^2(aK)}{(2\pi)^2}\right)$ and

$$\mathbb{F}_K = \left[\frac{\left(\frac{\partial \epsilon_{\alpha,surface}(K)}{\partial K} - \frac{\partial \Gamma_2(K)}{\partial K}\right)\left(\epsilon_{\alpha,surface}(K) - \Gamma_2(K)\right)}{\left\{\left(\epsilon_{\alpha,surface}(K) - \Gamma_2(K)\right)^2 + (CVb)^2\right\}}\right]. \quad (30)$$

It may be noted that we have put $\zeta\zeta' = +1$ in $F_{\alpha,\alpha,\zeta,\zeta'}(K,q) = \left(\frac{1}{2}\right)\left[1 + \zeta\zeta' \cos\theta_{\alpha,\alpha,K,q}\right]$ and wrote it as $(1 + (aq)\mathbb{F}_K)$ in Eq.(29), for we are considering the intra-band case presently. The integrals in (29) are trivial as $\frac{\partial n_{\zeta\alpha}(aK)}{\partial \mu}$ could be replaced by a delta function at T = 0K. In the high frequency limit Eq. (29) may further be written as

$$\left[\frac{(\hbar\omega)^3}{\left\{1+\frac{\pi V_0}{(\hbar\omega)}\int d\mathbf{K}\sum_{\alpha,\zeta}\frac{\partial n_{\zeta,\alpha}}{\partial\mu}\frac{\partial\epsilon_{\zeta,\alpha,surface}(a\mathbf{K})}{\partial(a\mathbf{K})}+\frac{\frac{1}{2}\pi V_0 (aq)^2}{(\hbar\omega)^3}\int d\mathbf{K}\sum_{\alpha,\zeta}\frac{\partial n_{\zeta,\alpha}}{\partial\mu}\frac{\partial\epsilon_{\zeta,\alpha,surface}(a\mathbf{K})}{\partial(a\mathbf{K})}\left(\frac{a|\mathbf{K}|}{\lambda_{\zeta,\alpha}}\right)^2\right\}}\right]$$

$$= (\hbar\omega)^2 (aq)\left\{\pi V_0 \int d\mathbf{K} \sum_{\alpha,\zeta}\frac{\partial n_{\zeta,\alpha}}{\partial\mu}\frac{\partial\epsilon_{\zeta,\alpha,surface}(a\mathbf{K})}{\partial(a\mathbf{K})}\mathbb{F}_K\right\}$$

$$+\frac{1}{2}\pi V_0 (aq)^3 \int d\mathbf{K} \sum_{\alpha,\zeta}\frac{\partial n_{\zeta,\alpha}}{\partial\mu}\frac{\partial\epsilon_{\zeta,\alpha,surface}(a\mathbf{K})}{\partial(a\mathbf{K})}\left(\frac{a|\mathbf{K}|}{\lambda_{\zeta,\alpha}}\right)^2 \mathbb{F}_K. \tag{31}$$

In the long wavelength limit, the second term in the right-hand-side of Equation (31) could be ignored compared to the first term. As a result this equation yields acoustic plasmons with group velocity proportional to the integral $\left\{\pi V_0 \int d\mathbf{K} \sum_{\alpha,\zeta}\frac{\partial n_{\zeta,\alpha}}{\partial\mu}\frac{\partial\epsilon_{\zeta,\alpha,surface}(a\mathbf{K})}{\partial(a\mathbf{K})}\mathbb{F}_K\right\}$. This velocity is several order of magnitude smaller than the speed of light. Thus, considering the intra-band transitions only, we have found possibility of only one collective mode exhibiting acoustic dispersion which could not be excited directly by light. It corresponds to charge plasmons. The linear behaviour of the dispersion implies that group and phase velocities are the same. So, signals can be transmitted undistorted along the surface. The finding has significant importance in extremely low loss communications.

The inter-band plasmons corresponding to a pair of spin-split bands will involve $\zeta\zeta' = -1$ in $F_{\alpha,\alpha,\zeta,\zeta'}(K,q) = \left(\frac{1}{2}\right)\left[1+\zeta\zeta'\cos\theta_{\alpha,\alpha,K,q}\right]$ which we write as $\left(\frac{1}{2}\right)(1-(aq)\mathbb{F}_K)$ below. Furthermore, the quantity $\varepsilon_r$ is the relative permittivity of the surrounding medium and it is positive (negative) for a dielectric (negative dielectric constant(NDC)) medium. We assume the surrounding medium to be medium with NDC. Thus, we may write $V_0 = -|V_0|$. Equation (31) in this case may be written as

$$\left[\frac{(\hbar\omega)^3}{\left\{1-\frac{\pi |V_0|}{(\hbar\omega)}\int d\mathbf{K}\sum_{\alpha,\zeta}\frac{\partial n_{\zeta,\alpha}}{\partial\mu}\frac{\partial\epsilon_{\zeta,\alpha,surface}(a\mathbf{K})}{\partial(a\mathbf{K})}-\frac{\frac{1}{2}\pi |V_0| (aq)^2}{(\hbar\omega)^3}\int d\mathbf{K}\sum_{\alpha,\zeta}\frac{\partial n_{\zeta,\alpha}}{\partial\mu}\frac{\partial\epsilon_{\zeta,\alpha,surface}(a\mathbf{K})}{\partial(a\mathbf{K})}\left(\frac{a|\mathbf{K}|}{\lambda_{\zeta,\alpha}}\right)^2\right\}}\right]$$

$$= (\hbar\omega)^2 (aq)\left\{\pi |V_0| \int d\mathbf{K} \sum_{\alpha,\zeta}\frac{\partial n_{\zeta,\alpha}}{\partial\mu}\frac{\partial\epsilon_{\zeta,\alpha,surface}(a\mathbf{K})}{\partial(a\mathbf{K})}\mathbb{F}_K\right\}$$

$$+\frac{1}{2}\pi |V_0| (aq)^3 \int d\mathbf{K} \sum_{\alpha,\zeta}\frac{\partial n_{\zeta,\alpha}}{\partial\mu}\frac{\partial\epsilon_{\zeta,\alpha,surface}(a\mathbf{K})}{\partial(a\mathbf{K})}\left(\frac{a|\mathbf{K}|}{\lambda_{\zeta,\alpha}}\right)^2 \mathbb{F}_K. \tag{32}$$

Since the second term in the right-hand-side is small compared to the first term, we expect to obtain acoustic plasmons even in this case with the same group velocity. Such possible acoustic plasmons, derived using the gapped surface state excitation spectrum, has the potential to settle the continuing row **[28,31]** over whether the surface metallicity with the insulating bulk or the bulk metallicity with gapped surface state excitation spectrum is dominant on the issue of the origin of observed anomalies at low temperatures in $SmB_6$. How? The former, i.e. the surface metallicity with the insulating bulk, is expected to give conventional $q^{1/2}$ plasmons, whereas the latter, i.e. the gapped surface state, is expected to yield plasmons with unconventional dispersion relation. Thus, the detection of unconventional (acoustic) plasmons experimentally establishes the dominance of the bulk metallicity. We now discuss the strong incarceration capability of these plasmon which is another notable outcome of the present work. This part of the paper leans heavily on the previous investigations **[36,37].**

### 3.3 Plasmon confinement measure

A figure-of-merit for the degree of Plasmon confinement to be introduced below involves dynamical optical conductivity $\sigma(q,\omega)$ of the system in an essential manner. The real part of the optical conductivity describes the dissipation of the electromagnetic energy in the medium, while the imaginary part describes screening of the applied field. One obtains the expression for the optical conductivity using the relation $\sigma(q,\omega) = \frac{e^2}{\hbar}(4i\,(a\,q)^{-2}\hbar\omega)\,\chi(aq,\omega)$. In view of (27), the imaginary part of the intra-band optical conductivity is proportional to $\frac{8\pi}{(aq)}$ $F_1$ where the function $F_1$ given by the expression

$$\sum_{K,\alpha,\zeta} \left[ \frac{\frac{\partial n_{\zeta,\alpha}}{\partial \mu} \frac{\partial \epsilon_{\zeta,\alpha,surface}(aK)}{\partial(aK)} F_{\alpha,\zeta}(K,q)}{\left\{1 - \left(\frac{a|q|\,a|k|}{\hbar\omega\,\lambda_{\zeta,\alpha}}\right)^2\right\}^{1/2}} \right]. \qquad (33)$$

In the first approximation, this is independent of $(aq)$ and $(\hbar\omega)$ in the long wavelength and high frequency limits. The real part, on the other hand, is given by $\frac{4\pi(\hbar\omega)^2}{(a\,q)^2}$ $F_2$ where $F_2$ is the sum

$$(\hbar\omega)^{-1} \sum_{K,\alpha,\zeta} (n_{\zeta,\alpha}(aK - aq) - n_{\zeta,\alpha}(aK)) F_{\alpha,\zeta}(K,q)\,\delta\!\left(\hbar\omega - \left\{\frac{(aq.aK)}{\lambda_{\zeta,\alpha}}\right\}\right). \qquad (34)$$

Since $[(n_{\zeta,\alpha}(aK-aq) - n_{\zeta,\alpha}(aK))\,F_{\alpha,\zeta}(K,q)]$ yields $[(aq)\frac{\partial n_{\zeta,\alpha}}{\partial \mu}\frac{\partial \epsilon_{\zeta,\alpha,surface}(aK)}{\partial(aK)} F_{\alpha,\zeta}(K,q)]$ and the derivative $\frac{\partial \epsilon_{\zeta,\alpha,surface}(aK)}{\partial(aK)}$, in view of (22), is

$$(t_{d1} + b^2 t_{f1})\,aK + \frac{v_F^*(\zeta)^2(aK)}{\sqrt{(CVb)^2 + \left(\frac{b^2}{2}\epsilon_f - 3t_{f1}\right)^2 + v_F^*(\zeta)^2(Ka)^2 + O((Ka)^4)}}, \qquad (35)$$

we notice that the term $[(n_{\zeta,\alpha}(aK-aq) - n_{\zeta,\alpha}(aK))\,F_{\alpha,\zeta}(K,q)]$ will involve a product $[(aq)(aK)]$. Since the delta function in (34) demands the product to be equal $[(\hbar\omega)\lambda_{\zeta,\alpha}]$, the term $(\hbar\omega)^{-1}$ before the sum in (34) will cancel out. We have, thus, $Re\sigma(q,\omega) = \frac{e^2}{\hbar}\frac{4\pi(\hbar\omega)^2}{(a\,q)^2}$ $F_2$ and $Im\sigma(q,\omega) = \frac{e^2}{\hbar}\frac{8\pi}{(aq)}$ $F_1$ where $F_2$ and $F_1$ do not depend upon $(aq)$ and $(\hbar\omega)$ in the first approximation. Now upon treating wave vector to be a complex variable : $q = q_1 + iq_2$ with $q_2 \neq 0$, we find that the ratio $\frac{q_1}{q_2} = \frac{\left[\frac{\partial}{\partial q_1}(q_1\, Im\sigma(q_1,\omega))\right]}{Re\sigma(q_1,\omega)}$. In the long wavelength and high frequency limits, this ratio is approximately $\frac{2(aq)}{(\hbar\omega)^2}(F_1/F_2) \ll 1$. We next need a figure-of-merit (FOM) which measures the confinement of the surface wave with respect to the corresponding vacuum wavelength.

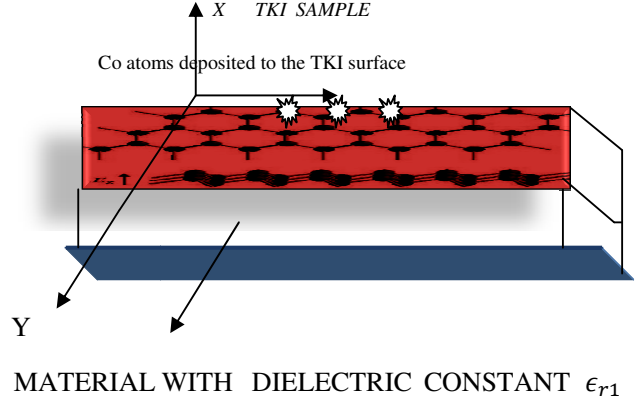

MATERIAL WITH DIELECTRIC CONSTANT $\epsilon_{r1}$

**Figure 1.** A schematic diagram of the TKI sample on a dielectric material.

In order to obtain a relevant FOM we consider a geometry shown in Figure 1 where TKI sample is surrounded with dielectrics of constants $\epsilon_{r1}$ and $\epsilon_{r2}$(air). We assume that the electric field has the form $E_z = Ae^{iqz-Qx}, E_y = 0, E_x = Be^{iqz-Qx}$, for $x > 0$, $E_z = Ce^{iqz+Q'x}, E_y = 0, E_x = De^{iqz+Q'x}$, for $x < 0$. After inserting this ansatz into Maxwells equations and matching the appropriate boundary conditions, in the non-retarded regime we obtain the dispersion relation for TM modes as $Q \approx Q' \approx q \approx \epsilon_0\epsilon\frac{2i\omega}{\sigma(\omega,q)}$, or $\epsilon + i\left(\frac{q}{2\omega\epsilon_0}\right)\sigma(q,\omega) = 0$ where the dielectric constant $\epsilon = (\epsilon_{r1} + \epsilon_{r2})/2$, $\epsilon_0$ is the vacuum permittivity, $\sigma(q,\omega)$ is the wavevector dependent optical conductivity, $q$ is a wave vector, and $\omega$ is the angular frequency of the incident monochromatic optical field. Incidentally, with our expression of $Im\sigma^{intraband}(q,\omega)$ above, upon using this standard equation, we obtain Eq. (41). This is a good consistency check; our result is consistent with Maxwells equations. Now an intuitive definition for a figure-of-merit for the degree of Plasmon confinement, on the other hand, is $FOM_{conf} = \lambda_0\, Im(Q) \approx \frac{Im(Q)}{Re(Q)} \approx \frac{q_2}{q_1}$ where $\lambda_0$ is the free-space wavelength, and $Im(Q)$ ($Re(Q)$) is the imaginary (real) part of the complex wave vector $Q$ perpendicular to the interface. This figure-of-merit measures the confinement of the surface wave with respect to the corresponding vacuum wavelength. Since we have found that $\frac{q_2}{q_1} \approx \frac{(\hbar\omega)^2}{2(aq)}(F_2/F_1) \gg 1$, the confinement measure of the surface wave is much larger than unity in the long wavelength and high frequency limits. Thus, apart from a linear dispersion, our theoretical investigation unveils strongly reduced plasmonic wavelengths compared to the free-space wavelength.

## 4. DISCUSSION AND CONCLUDING REMARKS

In the present communication we have started with PAM- a model for a generic TKI. The model itself and its extensions are still relevant for the theoretical condensed-matter physics. The examples of extension are those including the on-site interaction between *d*- and *f*-electrons and the nearest-neighbour interaction between *f*-electrons. A stronger version of this on-site interaction is found to destroy the Kondo state and narrow the intermediate valence regime **[5]**, while the latter allows to affect the stability of the magnetic ground state in the Kondo regime **[6]**. We also have introduced an extension to the model here by way of involving the magnetic exchange interaction. The interaction is introduced in the most direct route using only the spin degrees of freedom. We have demonstrated that the exchange field can be used to open a gap at the surface states with gapless Dirac dispersion. The field, in fact, is expected to play a bigger role, such as in the efficient tuning of the bulk band gap, in the context of the dielectric properties like the plasmon frequency, and so on. Furthermore, similar to what has been shown in the present work, the acoustic plasmons are a real possibility in other systems, such as graphene monolayer, as well. The dispersion and the damping of the sheet plasmon in a graphene monolayer grown on Pt(111) had been studied by using angle-resolved electron energy loss spectroscopy several years ago. The investigators **[38-40]** discovered that the dispersion relation of the plasmon mode confined in the graphene sheet was linear with considerably reduced wavelength compared to that of the free-space implying an improved field confinement. Those in ref.**[36]**, reported real-space imaging of acoustic THz plasmons in a graphene photodetector with split- gate architecture. For this purpose they introduced nanoscale-resolved THz photo-current near-field microscopy, where near-field excited graphene plasmons are detected thermoelectrically rather than optically. The photocurrent images also unveil strongly reduced plasmonic wavelengths, and a linear dispersion. These experimental techniques could also be used for the present TKI system with suitable modification to detect experimentally the possibility of acoustic plasmons. Looking backward, we observe there are many unsettled issues. For example, the problem of hybridization of plasmons with optical surface phonons that is likely to occur when TKI is surrounded by a material other than air has not been addressed. The system plasmons may suffer from high losses at infrared wavelengths which could be attributed to the presence of multiple damping pathways, such as collisions with impurities and phonons, as well as particle/hole generation via inter-band damping. We need to suggest the ways and means to curb these losses.

In conclusion, the plasmons in TKI have notable properties and offer promising prospects for plasmonic applications. Though immense progress has been made in plasmonics over the past several decades, never-the-less, we believe that our work casts new light onto the some of the important facts pertaining to the TKI plasmonics.There are many challenges in the processing of these exotic materials to use the metallic/insulating states in functional devices, and they present great opportunities for the materials science research communities.

## Appendix A

We follow the methodology pioneered by Kadanoff and Baym **[24]**, by and large, to accomplish the tasks of calculating the thermal average of the slave-boson Hamiltonian followed by the thermodynamic potential per unit volume( $\Omega_{sb}$). The method involves establishing a relation between this average and certain spectral-weight functions. Since $Lim_{\eta \to 1} V^{-1} \int \frac{d\eta}{\eta} \langle \aleph_{sb}(b,\lambda,\xi\,\eta) \rangle_\eta = \Omega_{sb}$ where the angular brackets $\langle \aleph_{sb}(b,\lambda,\xi\,\eta) \rangle_\eta$ denote thermodynamic average calculated with $\aleph_{sb}(b,\lambda,\xi\,\eta) = \eta \aleph_{sb}(b,\lambda,\xi)$, the task basically boils down to finding relations between $\langle \aleph_{sb}(b,\lambda,\xi\,\eta) \rangle_\eta$ and the spectral functions defined below. For this purpose, in this appendix, we set up equations of motion for the operators $d_{k\zeta}(t), s_{k,\zeta}(t)$, etc., where, in units such that $\hbar = 1$, $d_{k\zeta}(t) = \exp(i\aleph_{sb}(\eta)t)\, d_{k\zeta} \exp(-i\aleph_{sb}(\eta)t)$, and $s_{k,\zeta}(t) = \exp(i\aleph_{sb}(\eta)t)\, s_{k,\zeta} \exp(-i\aleph_{sb}(\eta)\,t)$. Here $\aleph_{sb}(\eta)$ means $\aleph_{sb}(b,\lambda,\xi\,\eta)$. As regards the equation of motion for the average $\langle \aleph_{sb}(\eta) \rangle_\eta$, we find

$$\langle \aleph_{sb}(\eta) \rangle_\eta = \frac{1}{2} \sum_{k,\zeta} [i\,\frac{\partial}{\partial t} - i\,\frac{\partial}{\partial t'}]\,\{\langle d^\dagger_{k,\zeta}(t')\, d_{k,\zeta}(t) \rangle_\eta + \langle bs^\dagger_{k,\zeta}(t')bs_{k,\zeta}(t) \rangle_\eta\} + \lambda\eta N_s\,(b^2-1), \text{ (A.1)}$$

where $\langle d^\dagger_{k,\zeta}(t')\, d_{k,\zeta}(t) \rangle_\eta$ etc., are expressible in terms of the spectral-weight functions $A^{(\eta)}_{sb}(k\zeta,\omega)$ and $B^{(\eta)}_{sb}(k\zeta,\omega)$ given by

$$A_{sb}^{(\eta)}(k\zeta,\omega) = i[G_{sb}^{(\eta)}(k\zeta,k\zeta,\omega_n)\big|_{i\omega_n=\omega+i0^+} - G_{sb}^{(\eta)}(k\zeta,k\zeta,\omega_n)\big|_{i\omega_n=\omega-i0^+}],$$
$$B_{sb}^{(\eta)}(k\zeta,\omega) = i[F_{sb}^{(\eta)}(k\zeta,k\zeta,\omega_n)\big|_{i\omega_n=\omega+i0^+} - F_{sb}^{(\eta)}(k\zeta,k\zeta,\omega_n)\big|_{i\omega_n=\omega-i0^+}], \quad (A.2)$$

where $G_{sb}^{(\eta)}(k\zeta,k'\zeta',\omega_n)$ and $F_{sb}^{(\eta)}(k\zeta,k'\zeta',\omega_n)$, respectively, are the Fourier coefficients of the temperature Green's functions $G_{sb}^{(\eta)}(k\tau\zeta,k'\tau'\zeta') = -\langle T_\tau\{d_{k,\zeta}(\tau)d^\dagger_{k',\zeta'}(\tau')\}\rangle_\eta$, and $F_{sb}^{(\eta)}(k\tau\zeta,k'\tau'\zeta') = -\langle T_\tau\{bs_{k,\zeta}(\tau)bs^\dagger_{k',\zeta'}(\tau')\}\rangle_\eta$ where the time evolution an operator $O$ is given by $O(\tau)=\exp(\aleph_{sb}(\eta)\tau) O \exp(-\aleph_{sb}(\eta)\tau)$, $T_\tau$ is the imaginary time-ordering operator which arranges other operators from right to left in the ascending order of time $\tau$, and $\omega_n = \frac{(2n+1)\pi}{\beta}, n = 0, \pm 1, \pm 2, \ldots$ The LR of $A_{sb}^{(\eta)}(k\zeta,\omega)$ and $B_{sb}^{(\eta)}(k\zeta,\omega)$ are easy to write. With the help of these representations one obtains $A_{sb}^{(\eta)}(k\zeta,\omega) = -2(e^{\beta\omega}+1)Im \, \mathring{A}_{sb}^{(\eta)}(k\zeta,\omega)$ and $B_{sb}^{(\eta)}(k\zeta,\omega) = -2(e^{\beta\omega}+1)Im \, \mathring{B}_{sb}^{(\eta)}(k\zeta,\omega)$ where

$$\mathring{A}_{sb}^{(\eta)}(k\zeta,\omega) = -i\int dt \int dt' \, e^{i\omega(t-t')} \langle d^\dagger_{k,\zeta}(t') d_{k,\zeta}(t)\rangle_\eta \, \theta(t-t'),$$
$$\mathring{B}_{sb}^{(\eta)}(k\zeta,\omega) = -i\int dt \int dt' \, e^{i\omega(t-t')} \langle bs^\dagger_{k,\zeta}(t') bs_{k,\zeta}(t)\rangle_\eta \, \theta(t-t'). \quad (A.3)$$

In view of the LRs, it is easy to see that $\langle d^\dagger_{k,\zeta}(t') d_{k,\zeta}(t)\rangle_\eta = \int_{-\infty}^{+\infty} \frac{d\omega}{2\pi} \frac{A_{sb}^{(\eta)}(k\zeta,\omega)}{(e^{\beta\omega}+1)} e^{-i\omega(t-t')}$, and $\langle bs^\dagger_{k,\zeta}(t') bs_{k,\zeta}(t)\rangle_\eta = \int_{-\infty}^{+\infty} \frac{d\omega}{2\pi} \frac{B_{sb}^{(\eta)}(k\zeta,\omega)}{(e^{\beta\omega}+1)} e^{-i\omega(t-t')}$. Upon substituting these results in $\langle \aleph_{sb}(\eta)\rangle_\eta$ one obtains the relation sought for:

$$\langle \aleph_{sb}(\eta)\rangle_\eta = \frac{1}{2}\sum_{k,\zeta}\int_{-\infty}^{+\infty}\frac{d\omega}{2\pi}\frac{1}{(e^{\beta\omega}+1)}\left[A_{sb}^{(\eta)}(k\zeta,\omega) + B_{sb}^{(\eta)}(k\zeta,\omega)\right] + \lambda\eta \, N_s \, (b^2-1). \quad (A.4)$$

The weight functions in (A.4) can now be calculated setting up equations of motion for the temperature Green's functions $G_{sb}^{(\eta)}(k\tau\zeta,k'\tau'\zeta')$, and $F_{sb}^{(\eta)}(k\tau\zeta,k'\tau'\zeta')$ and transforming these equations to the ones for the corresponding Fourier coefficients. The latter constitute a system of homogeneous, linear equations. For the Fourier coefficients, one obtains

$$G_{sb}^{(\eta)}(k\uparrow,k\uparrow,\omega_n) = \frac{u_{k,+}^{(-)2}}{i\omega_n-\epsilon_-^{(-)}(\eta,k)} + \frac{u_{k,-}^{(-)2}}{i\omega_n-\epsilon_+^{(-)}(\eta,k)}, \quad G_{sb}^{(\eta)}(k\downarrow,k\downarrow,\omega_n) = \frac{u_{k,+}^{(+)2}}{i\omega_n-\epsilon_-^{(+)}(\eta,k)} + \frac{u_{k,-}^{(+)2}}{i\omega_n-\epsilon_+^{(+)}(\eta,k)},$$

$$F_{sb}^{(\eta)}(k\uparrow,k\uparrow,\omega_n) = \frac{u_{k,+}^{(+)2}-v_k^{(+)2}}{i\omega_n-\epsilon_-^{(+)}(\eta,k)} + \frac{u_{k,-}^{(+)2}+v_k^{(-)2}}{i\omega_n-\epsilon_+^{(+)}(\eta,k)} + \frac{v_k^{(+)2}}{i\omega_n-\epsilon_-^{(-)}(\eta,k)} + \frac{-v_k^{(-)2}}{i\omega_n-\epsilon_-^{(-)}(\eta,k)},$$

$$F_{sb}^{(\eta)}(k\downarrow,k\downarrow,\omega_n) = \frac{-v_k^{(+)2}}{i\omega_n-\epsilon_-^{(+)}(\eta,k)} + \frac{v_k^{(-)2}}{i\omega_n-\epsilon_+^{(+)}(\eta,k)} + \frac{u_{k,-}^{(+)2}+v_k^{(+)2}}{i\omega_n-\epsilon_+^{(-)}(\eta,k)} + \frac{u_{k,+}^{(+)2}-v_k^{(-)2}}{i\omega_n-\epsilon_-^{(-)}(\eta,k)}, \quad (A.5)$$

$$\epsilon_\alpha^{(\zeta)}(\eta,k) = -\frac{\eta(\varepsilon_d+\varepsilon_f+\zeta M)}{2} + \eta\alpha\sqrt{\frac{(\varepsilon_d-\varepsilon_f+\zeta M)^2}{4} + 4V^2b^2(s_x^2+s_y^2+s_z^2)},$$

$$\varepsilon_d(k) = (\mu+\xi+\varepsilon_k^d), \quad \varepsilon_f(k) = (\mu-\xi+b^2\varepsilon_k^f-\lambda),$$

$$\epsilon_k^d = [2t_{d1}c_1(k) + 4t_{d2}c_2(k) + 8t_{d3}c_3(k)],$$

$$\epsilon_k^f = [-\epsilon_f + 2t_{f1}\, c_1(k) + 4t_{f2}\, c_2(k) + 8t_{f3}\, c_3(k)]. \tag{A.6}$$

The coherence factors $(u_{k,+}^{(\zeta)^2}, u_{k,-}^{(\zeta)^2})$ are given by

$$u_{k,\pm}^{(\zeta)^2} = \tfrac{1}{2}[1 \pm \frac{(\varepsilon_d - \varepsilon_f + \zeta M)}{2\{\sqrt{\frac{(\varepsilon_d - \varepsilon_f + \zeta M)^2}{4} + 4V^2 b^2 (s_x^2 + s_y^2 + s_z^2)}\}}]. \tag{A.7}$$

$$v_k^{(\sigma)^2} = \frac{-2V^2 b^2 s_z^2}{\sqrt{\frac{(\varepsilon_d - \varepsilon_f - M)^2}{4} + 4V^2 b^2 (s_x^2 + s_y^2 + s_z^2)} \, [\frac{(\varepsilon_d - \varepsilon_f - M)}{2} + \sigma \sqrt{\frac{(\varepsilon_d - \varepsilon_f - M)^2}{4} + 4V^2 b^2 (s_x^2 + s_y^2 + s_z^2)}]}, \tag{A.8}$$

where $\alpha = 1\, (-1)$ for upper band (lower band), $\zeta = \pm 1$ labels the eigenstates ($\uparrow, \downarrow$) of $\zeta_z$, and $\sigma = \pm 1$. Upon using the Sokhotski-Plemelj identity $\{x \pm i\eta\}^{-1} = P(x^{-1}) \mp i\pi\delta(x)$, in view of (A.2) and (A.5), we finally find that the spectral weight in (A.4) and $\langle \aleph_{sb} \rangle$ are given by a bunch of delta functions:

$$\langle \aleph_{sb} \rangle = \sum_k \int d\omega (e^{\beta\omega} + 1)^{-1} \left[ (u_{k,+}^{(+)^2} - v_k^{(+)^2}) \delta(\omega - \eta\, \in_-^{(+)}) + (u_{k,-}^{(+)^2} + v_k^{(-)^2}) \delta(\omega - \eta\, \in_+^{(+)}) \right.$$

$$\left. + \{\frac{u_{k,-}^{(-)^2} + u_{k,-}^{(+)^2}}{2} + v_k^{(+)^2}\} \delta(\omega - \eta\, \in_+^{(-)}) + \{\frac{u_{k,-}^{(+)^2} + u_{k,+}^{(-)^2}}{2} - v_k^{(+)^2}\} \delta(\omega - \eta\, \in_-^{(-)}) \right]. \tag{A.9}$$

This is easy to integrate leading eventually to an expression for the thermodynamic potential.

**Appendix B**

The scheme to calculate the averages $\langle d_{k,\zeta}^\dagger d_{k,\zeta} \rangle, \langle bs_{k,\zeta}^\dagger bs_{k,\zeta} \rangle$, etc. have been shown in this appendix. For this purpose, we proceed with finite-temperature formalism. Since the Hamiltonian is completely diagonal one can write down easily the equations for the operators $\{d_{k,\zeta}(\tau), s_{k,\zeta}(\tau)\}$, where the time evolution an operator $O$ is given by $O(\tau) = \exp(\aleph\tau)\, O\, \exp(-\aleph\tau)$, to ensure that the thermal averages in the equations above are determined in a self-consistent manner. The Green's functions $G_{sb}(k\zeta, k\zeta, \tau) = -\langle T_\tau \{d_{k,\zeta}(\tau) d^\dagger_{k,\zeta}(0)\} \rangle$, $F_{sb}(k\zeta, k\zeta, \tau) = -b^2 \langle T_\tau \{s_{k,\zeta}(\tau) d^\dagger_{k,\zeta}(0)\} \rangle$, etc., where $T_\tau$ is the time-ordering operator which arranges other operators from right to left in the ascending order of imaginary time $\tau$, are of primary interest. We find

$$G_{sb}(k\uparrow, k\uparrow, \tau \to 0^+) = u_{k,+}^{(-)^2} (e^{\beta(\in_-^{(-)}(k) - \mu)} + 1)^{-1} + u_{k,-}^{(-)^2} (e^{\beta(\in_+^{(-)}(k) - \mu)} + 1)^{-1}, \tag{B1}$$

$$G_{sb}(k\downarrow, k\downarrow, \tau \to 0^+) = u_{k,+}^{(+)^2} (e^{\beta(\in_-^{(+)}(k) - \mu)} + 1)^{-1} + u_{k,-}^{(+)^2} (e^{\beta(\in_+^{(+)}(k) - \mu)} + 1)^{-1}, \tag{B2}$$

$$F_{sb}(k\uparrow, k\uparrow, \tau \to 0^+) = (u_{k,+}^{(+)^2} - v_k^{(+)^2})(e^{\beta(\in_-^{(+)}(k) - \mu)} + 1)^{-1} + (u_{k,-}^{(+)^2} + v_k^{(-)^2})(e^{\beta(\in_+^{(+)}(k) - \mu)} + 1)^{-1}$$

$$+ v_k^{(+)^2} (e^{\beta(\in_+^{(-)}(k) - \mu)} + 1)^{-1} - v_k^{(-)^2} (e^{\beta(\in_-^{(-)}(k) - \mu)} + 1)^{-1}, \tag{B3}$$

$$F_{sb}(k\downarrow, k\downarrow, \tau \to 0^+) = (u_{k,-}^{(+)^2} + v_k^{(+)^2})(e^{\beta(\in_+^{(-)}(k) - \mu)} + 1)^{-1} + (u_{k,+}^{(+)^2} - v_k^{(-)^2})(e^{\beta(\in_-^{(-)}(k) - \mu)} + 1)^{-1}$$

$$- v_k^{(+)^2} (e^{\beta(\in_-^{(+)}(k) - \mu)} + 1)^{-1} + v_k^{(-)^2} (e^{\beta(\in_+^{(+)}(k) - \mu)} + 1)^{-1}. \tag{B4}$$

,

For the averages $\langle d^\dagger_{k\uparrow} bs_{k,\downarrow}\rangle$ and $\langle bs^\dagger_{k,\uparrow} d_{k\downarrow}\rangle$, respectively, we obtain $\frac{V(s_x+i s_y)}{\varepsilon_-(k,b,\lambda,\xi)}\left[(e^{\beta(\epsilon^{(-)}_-(k)-\mu)} + 1)^{-1} - (e^{\beta(\epsilon^{(-)}_+(k)-\mu)} + 1)^{-1}\right]$ and $\frac{V(s_x+i s_y)}{\varepsilon_+(k,b,\lambda,\xi)}\left[(e^{\beta(\epsilon^{(+)}_-(k)-\mu)} + 1)^{-1} - (e^{\beta(\epsilon^{(+)}_+(k)-\mu)} + 1)^{-1}\right]$. These averages involving hybridization parameter *V* are ultimate signature of the Kondo insulating state, where there is precisely one conduction electron paired with an impurity spin. In the zero-temperature limit the Fermi functions $(e^{\beta(\epsilon(k)-\mu)} + 1)^{-1}$ will be replaced by the Heaviside step function $\theta(\mu-\epsilon(k))$. Here

$$u^{(\zeta)^2}_{k,\pm} = \frac{1}{2}\left[1 \pm \frac{(2\xi+\epsilon^d_k - b^2\epsilon^f_k+\lambda+\zeta M)}{2\left\{\sqrt{\frac{(2\xi+\epsilon^d_k-b^2\epsilon^f_k+\lambda+\zeta M)^2}{4} + 4V^2 b^2(s_x^2+s_y^2+s_z^2)}\right\}}\right], \tag{B5}$$

$$v^{(\sigma)^2}_k = \frac{-2V^2 b^2 s_z^2}{\sqrt{\frac{(2\xi+\epsilon^d_k-b^2\epsilon^f_k+\lambda-M)^2}{4}+4V^2 b^2(s_x^2+s_y^2+s_z^2)}\left[\frac{(2\xi+\epsilon^d_k-b^2\epsilon^f_k+\lambda-M)}{2}+\sigma\sqrt{\frac{(2\xi+\epsilon^d_k-b^2\epsilon^f_k+\lambda-M)^2}{4}+4V^2 b^2(s_x^2+s_y^2+s_z^2)}\right]}, \tag{B6}$$

$$\varepsilon_-(k,b,\lambda,\xi) = \sqrt{\frac{(2\xi+\epsilon^d_k-b^2\epsilon^f_k+\lambda-M)^2}{4} + 4V^2 b^2(s_x^2+s_y^2+s_z^2)}, \tag{B7}$$

$$\varepsilon_+(k,b,\lambda,\xi) = \sqrt{\frac{(2\xi+\epsilon^d_k-b^2\epsilon^f_k+\lambda+M)^2}{4} + 4V^2 b^2(s_x^2+s_y^2+s_z^2)}, \tag{B8}$$

$$\epsilon^{(\zeta)}_\alpha(k) = -\frac{(\epsilon^d_k+b^2\epsilon^f_k-\lambda+\zeta M)}{2} + \alpha\sqrt{\frac{(2\xi+\epsilon^d_k-b^2\epsilon^f_k+\lambda+\zeta M)^2}{4}+4V^2 b^2(s_x^2+s_y^2+s_z^2)}, \tag{B9}$$

$$\epsilon^d_k = [2t_{d1} c_1(k) + 4t_{d2} c_2(k) + 8t_{d3} c_3(k)],$$

$$\epsilon^f_k = [-\epsilon_f + 2t_{f1} c_1(k) + 4t_{f2} c_2(k) + 8t_{f3} c_3(k)], \tag{B10}$$

and $\alpha = 1\ (-1)$ for upper band (lower band), $\zeta = \pm 1$ labels the eigenstates ($\uparrow,\downarrow$) of $\zeta_z$.

The corresponding author states that there is no conflict of interest.